\documentclass[apj]{emulateapj}
\usepackage{apjfonts}
\usepackage{graphicx}  

\begin{document}

\title{The Dynamical Distinction between Elliptical and Lenticular Galaxies in Distant Clusters:
Further Evidence for the Recent Origin of S0 Galaxies}
\author{Sean M. Moran\altaffilmark{1},  Boon Liang
  Loh\altaffilmark{2}, Richard S. Ellis,\altaffilmark{1}, Tommaso
  Treu\altaffilmark{3},  Kevin Bundy\altaffilmark{4}, Lauren A. MacArthur\altaffilmark{1}}
\altaffiltext{1}{California Institute of Technology, Department of Astronomy, MS 105-24, Pasadena, CA
  91125, USA email: smm@astro.caltech.edu, rse@astro.caltech.edu, lam@astro.caltech.edu}
\altaffiltext{2}{National University of Singapore, Department of
  Physics, Singapore 119077, email:u0301419@nus.edu.sg} 
\altaffiltext{3}{University of California, Santa Barbara, Department of Physics  CA 93106, email: tt@physics.ucsb.edu}
 \altaffiltext{4}{University of Toronto, Department of  Astronomy, Toronto, Ontario M5S 3H4, email: bundy@astro.utoronto.ca}

\begin{abstract}
We examine resolved spectroscopic data obtained with the Keck II telescope for 44 
spheroidal galaxies in the fields of two rich clusters, Cl0024+16 ($z$=0.40) and 
MS0451-03 ($z$=0.54), and contrast this with similar data for 23 galaxies within the 
redshift interval  0.3$<z<$0.65 in the GOODS northern field. For each galaxy we examine 
the case for systemic rotation, derive central stellar velocity dispersions $\sigma$ and 
photometric ellipticities, $\epsilon$. Using morphological classifications
obtained via Hubble Space Telescope imaging as the basis, we explore the 
utility of our kinematic quantities in distinguishing between pressure-supported ellipticals
and rotationally-supported lenticulars (S0s). We demonstrate the reliability of
using the $v/(1-\epsilon)$ vs $\sigma$ and $v/\sigma$ vs $\epsilon$ distributions
as discriminators, finding that the two criteria correctly identify
$63\%\pm3\%$ and 80$\%\pm2\%$ of S0s at $z\sim0.5$, respectively, along with $76\%\pm^8_3\%$
and 79$\%\pm2\%$ of ellipticals. We test these diagnostics using equivalent local data in
the Coma cluster, and find that the diagnostics are similarly accurate at $z=0$. Our measured accuracies 
are comparable to the accuracy of visual classification of morphologies, but avoid the band-shifting and
surface brightness effects that hinder visual classification at high
redshifts. As an example application of our kinematic discriminators, 
we then examine the morphology-density relation for elliptical 
and S0 galaxies separately at $z\sim$0.5. We confirm, from kinematic 
data alone, the recent growth of rotationally-supported spheroidals. We discuss the feasibility of 
extending the method to a more comprehensive study of cluster and field galaxies to 
$z\simeq$1, in order to verify in detail the recent density-dependent growth of S0
galaxies.

\end{abstract}


\section{INTRODUCTION}

Understanding the origin of the Hubble sequence remains a fundamental goal 
in extragalactic astronomy. The lasting utility of Hubble's classification scheme lies 
in its ability to distinguish between both the {\it dynamics} and {\it stellar populations} of 
disk and spheroidal galaxies. However, despite considerable progress in
unraveling the time evolution of elliptical and spiral galaxies (see Ellis 2007
for a recent review), there is still some disagreement concerning the origin
of lenticular (or S0) galaxies - a hybrid class with kinematic characteristics
of disk galaxies but whose present-day stellar populations resemble those 
seen in ellipticals (Es). Were S0s created {\it ab initio} or do they represent 
spirals whose gas supply was more recently exhausted? 
Understanding the origin of  this intriguing population is important in 
resolving the extent to which galaxies are morphologically influenced by 
their environment.

The idea that infalling cluster spirals have 
somehow been converted into S0s has received much support because of 
the presence of a local  {\it morphology-density relation} \citep{d80} and
evidence of its evolution \citep{d97, tt03}.
However, detailed studies of local S0s have failed to resolve the
question of whether they are faded remnants of early spirals 
(c.f. Poggianti et al. 2001 {\it vs.} Burstein et al. 2005), or if
they instead have similar formation histories to ellipticals, but
with different bulge-to-disk ratios. 
The most direct approach to resolving this debate would be to track directly
the evolution in the S0 and elliptical fractions with lookback time,
thus tracing the formation histories of the two classes independently.

\begin{figure*}[th]
\includegraphics[width=2\columnwidth]{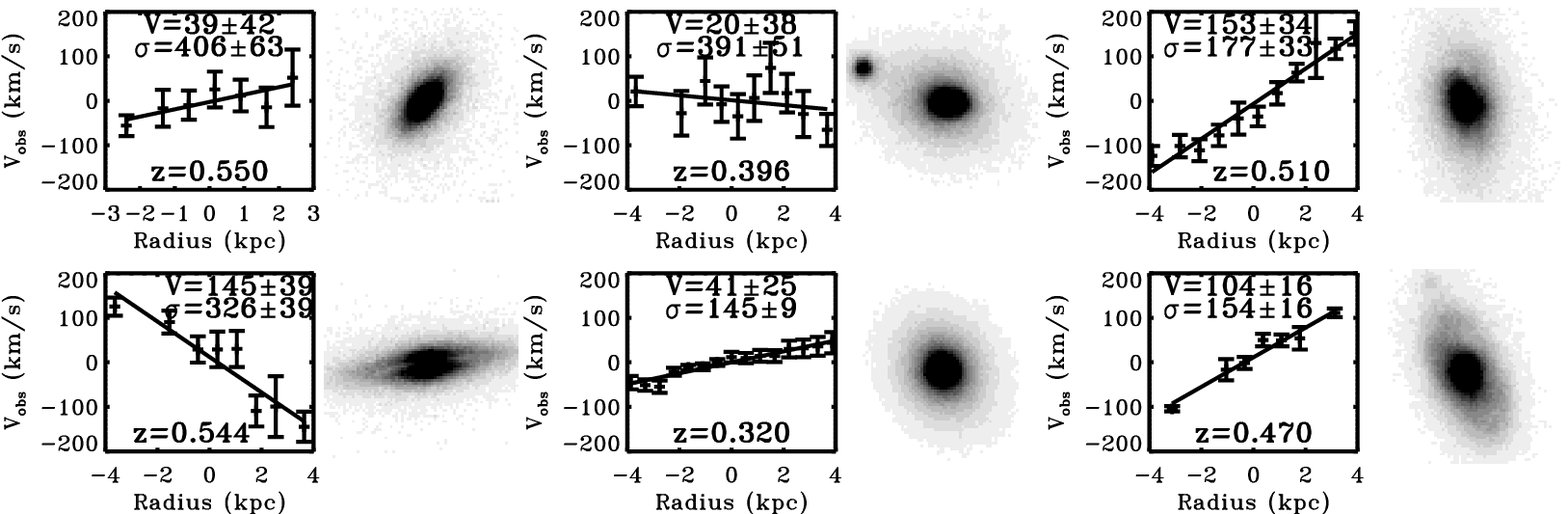}
\caption{
Selection of HST $F814W$ images and rotation curves
for both field and cluster spheroidals. From top left to bottom right:
two cluster Ellipticals, two cluster S0s, one field E and one field S0.
In each case the redshift $z$, rotational velocity, $v$, and stellar
velocity dispersion, $\sigma$, are listed with their errors.}
\label{fig:rotation_stamps}
\end{figure*}

Considerable progress has been made in tracking the
evolving fraction of spheroidals, $f_{E+S0}$, as a function 
of environmental density $\Sigma$ \citep{smith05,postman05,capak07}. 
Although the morphology-density relation was apparently
in place at $z\simeq$1, it has evolved quite substantially at later times,
mostly in regions of high projected density. One suggestion is that only 
ellipticals were present in abundance at z$\simeq$1 ($f_{S0}<$0.1), with 
subsequent growth in $f_{E+S0}$ arising primarily via a density-dependent transformation of 
spirals into S0s \citep[c.f.][]{smith05}. This simple hypothesis could be tested 
by separating S0s from ellipticals, so that their fraction, $f_{S0}$, could be 
determined independently of that of ellipticals as a function of both 
$\Sigma$ and $z$. If spiral transformations occurred, there 
should be fewer S0s in all over-dense environments at z$\simeq$1. 

Morphologically distinguishing distant S0s from their elliptical counterparts via 
Hubble Space Telescope (HST) imaging has proved difficult because of surface 
brightness dimming, loss of resolution, band-shifting and inclination effects, 
each of which might be redshift-dependent thereby introducing
biases. \citet{postman05} 
attempted to measure the S0 fraction morphologically at $z\simeq$1, 
but the scatter in their S0 classifications, as determined in various ways, 
implies uncertainties of $\delta f_{S0}>$0.15. Although they find $f_{S0}$ 
drops with redshift, the residual fraction at $z\simeq$1 could be  0--30\%, 
consistent with a range of hypotheses. Recently, \citet{desai07} have
found similarly low S0 fractions for clusters across the range
$0.5<z<0.8$, but uncertainties and cluster to cluster variation remain
equivalent to that of \citet{postman05}.

The purpose of this paper is to explore the use of {\it kinematic data} to 
improve the separation of distant Es and S0s. Luminous ellipticals are 
pressure-supported whereas S0s have circularly rotating disks. The 
necessary ingredients to enable this distinction are the rotational 
velocity $v$, the central stellar velocity dispersion, $\sigma$, and the photometric
ellipticity, $\epsilon$. The combination of these three quantities has
been famously used locally to demonstrate that, for a given $\epsilon$, 
ellipticals have $v\,/\,\sigma$ ratios less than that for a 
rotationally-supported spheroid \citep{binney82}. Tying 
morphological classifications to physical quantities (pressure, mass and angular momentum) 
across all epochs should reduce redshift-dependent biases and facilitate comparison 
to numerical models which incorporate environmental effects 
\citep[e.g.][]{del06}.

As a result of a campaign in two intermediate redshift ($z\simeq$0.5) clusters, 
we have secured Keck spectroscopic data for a large sample of galaxies 
spanning a wide range of environments. Morphological classifications are 
available as a training set in these samples. We utilize these data to test the
conjecture that a kinematic classifier can be reliably used to isolate
S0s from ellipticals at $z\simeq$0.5. After verifying that the classifiers robustly
isolate S0s from ellipticals in a local cluster sample, we then use them
to determine the S0 fraction as a function of environmental 
density, $\Sigma$, at this epoch. We show this is a promising approach and 
discuss the prospects for extending it to higher redshift samples so that the S0 
fraction might be completely mapped as a function of $\Sigma$ since $z\simeq$1. 
Throughout this paper we adopt a standard cosmology with $H_0=70.0$
km~s$^{-1}$ Mpc$^{-1}$, $\Omega_m=0.3$, and $\Omega_\Lambda=0.7$.

\section{OBSERVATIONS}

The primary dataset for this paper is a comprehensive Keck spectroscopic
and HST imaging (F814W) survey of morphologically-selected 
spheroidals in two clusters, Cl0024+16 ($z$=0.40) and MS0451-03 ($z$=0.54) 
\citep{tt03,moran05}. We contrast this with equivalent 
data taken in the northern GOODS field \citep{tt05a,tt05b}. The above cited
articles give full details of the morphological selection and spectroscopic
campaigns. Here we give a brief synopsis of the salient points.

The cluster data comprises 44 member spheroidals spanning
a wide range of cluster-centric radius to rest-frame $M_V=$-19.9,
corresponding approximately to the luminosity limit adopted by
\citet{d97} for the local and $z\sim0.5$ morphology-density relations. 
To derive rest-frame absolute magnitudes ($M_V$), we apply
k--corrections to the observed F814W photometry as in \citep{moran07}.

Morphological classification is discussed by \citet{tt03} for Cl0024+16
and in Moran et al. (2007b, in preparation) for MS0451-03, and has been
deemed reliable to at least $M_V=-19.5$. In cases where the
distinction between Es and S0s was ambiguous, we assigned the class E/S0.
To the adopted magnitude limit, successful morphological distinction
between Es and S0s was possible in 95\% of cases, with only 5\%
classed E/S0. The original $z'<$22.5 
GOODS-N sample was classified in the same visual manner by one of us (RSE), 
albeit from deeper HST data \citep{bundy05,tt05a,tt05b}. For this study, 
the sample has been restricted to the redshift range 0.3$<z<$0.65 and cut at 
$M_V>$-19.9 to provide a comparison sample of 23 field spheroidals.

All spectroscopic data were taken with the DEIMOS spectrograph
 \citep{deimos} on Keck II. The cluster sample was observed with a 900 line
grating (Cl0024) or 600 line grating (MS0451) in the 4500--8000 \AA\
region offering a resolution of $\sigma\simeq$30--50 km s$^{-1}$. Typical
exposure times were 2.5 hrs in Cl0024 and 4~hrs in MS0451. 
The field survey was designed to sample
higher redshift spheroidals in the OH forest and a 1200 line grating was used
in the 6700--9300 \AA\ region providing a resolution of $\sigma\simeq$20--30 km 
s$^{-1}$; exposure times for the brighter objects considered here were
typically 4 hrs.
In planning the spectroscopic observations, it was not always possible to align the slit 
along the major axis. We discarded galaxies where the orientational mismatch was
greater than 45 degrees, and for those with a smaller misalignment we
apply a correction to the measured velocities in our analysis below (\S3).

\section{KINEMATIC MEASUREMENTS}

All spectra were reduced using the DEEP2 pipeline \citep{davis03} using
procedures described by \citet{moran05} and \citet{tt05b}. 
Central stellar velocity dispersions for both field and cluster
spheroidals come from these analyses. Comprehensive tests were undertaken to
evaluate both instrumental and algorithmic uncertainties, with the
results indicating that the dispersions are accurate to $<10\%$.

Resolved (rotational) velocity data was secured via an extension of the
cross-correlation technique used to determine velocity dispersions.
Our resolved 2D spectra consist of a number of individual spectra, one
for each pixel across the spatial dimension of the slit.
To ensure reliable measurements, every 2D spectrum was re-binned along 
the spatial dimension to secure a minimum average signal to noise ($S/N$) of 3
($\mbox{\AA}^{-1}$) per bin. Galaxies with less than three spatial bins of 
sufficient $S/N$ were removed; for our typical seeing of $0\farcs7$, this
defines the minimum extent where a velocity gradient can be resolved.

For each of these spatial bins (with one spectrum for each), 
we independently determine the best-fitting velocity relative to the 
galaxy's mean redshift, via a cross-correlation with spectra of 
eight template G/K stars, shifted to the redshift of the galaxy.
The resulting rotation curves were averaged across all eight template fits. 
We then perform a linear least squares fit to the data points, and
define the rotational velocity, $v$, as half the velocity 
range of the fitted line from end to end of the measured curve. A simple
fit is justified because, unlike emission-line data, our absorption line curves 
rarely extend far enough to reach the characteristic `turnover',
$v_{rot}$, of a disk rotation curve.  The 
uncertainty in $v$ was taken to be the larger of the RMS deviation of each 
data point from the fitted line and the RMS deviation of each velocity measurement 
across template fits.

The allowed misalignment between the galaxy major axis and the
spectroscopic slit can reduce the measured $v$ by up to $\sim40\%$. 
We make a first-order correction to the measured velocities by
dividing by a factor of $\cos(\delta)$, where $\delta$ is the angle
between the galaxy major axis and the long axis of the spectroscopic slit. Such a
correction is generally only valid under the assumption that each
galaxy is a thin disk \citep{kapferer06}, yet our sample contains both disk and
spheroidal galaxies. In practice, however, the magnitude of the
correction is negligible for slowly rotating or non-rotating
spheroidals, and so only truly-rotating disks will have their measured
velocities altered significantly.

Figure~1 displays a sample of images alongside the respective rotation curves.
To measure photometric parameters of each galaxy, we make use of the
GALFIT software \citep{peng02}, fitting each galaxy to a 2D model
following the S\' ersic profile (see, e.g. Moran et al. 2005). In each
model, the S\' ersic parameter {\it n}, position angle (PA),
ellipticity ($\epsilon$), and  effective radius ($R_e$) are allowed to
vary freely, and the each model is convolved with a PSF derived from 
a star observed in the same {\it HST} mosaic.
For the analysis discussed below, we adopt the fitted values
of $\epsilon$ and $R_e$. Typically our measured absorption line
rotation curves extend to about $0.25 R_e$. Measured rotation
velocities, velocity dispersions, ellipticities and other basic info
for all 67 galaxies in our combined cluster and field sample are listed in Table~1.

\section{KINEMATIC CLASSIFIERS}

In identifying pressure-supported ellipticals from kinematic parameters, we began
by considering the $v/\,\sigma - \epsilon$ distribution originally proposed by 
\citet{binney82}, where $v/\sigma^*\approx v/\sigma\sqrt{
(1-\epsilon)/\epsilon} > 1$ indicates a rotationally
supported spheroid.
Unfortunately, our rotation curves do not extend to large enough
radius to reliably measure the maximum rotation, $v_{rot}$. Also, in
the case of S0s, the central velocity dispersion arises partly from the prominent spheroidal
component, rather than the disk. This mixing of bulge and disk information illuminates a more
fundamental problem: since the \citep{binney82} criterion
is meant to measure the degree of rotational support for a single
spheroid, it is not a sensible test to apply to a two-component
bulge-plus-disk S0. Consequently, we deemed the $v/\,\sigma^*$ ratio to
be ineffective for our purposes. 

\begin{figure}
\includegraphics[width=\columnwidth]{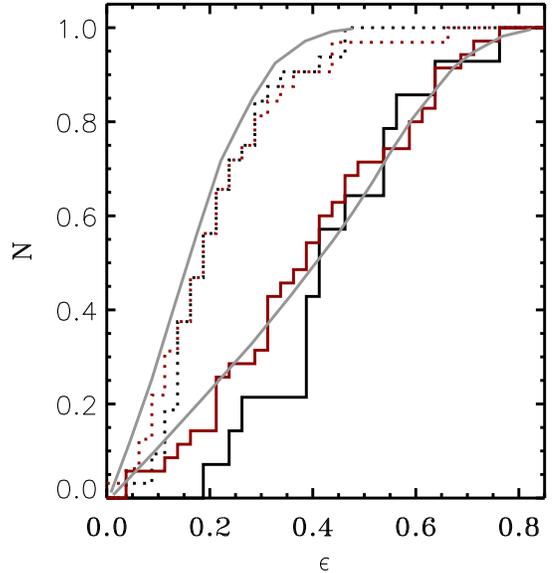}
\caption{\label{ellip} Cumulative distribution of ellipticities for
  galaxies in our $z\sim0.5$ (red) and local (black) samples. Dotted
  histograms indicate visually-classified ellipticals, and solid
  histograms denote S0s. The grey curves overplotted are the
  best-fitting Gaussian distributions of galaxy ellipticities from
  \citet{jf94}. At $z\sim0.5$, both Es and S0s are consistent with the
  model curves, indicating that the distributions are unbiased. The
  local sample exhibits the undercounting of low-ellipticity S0s
  discussed in \citet{jf94}.
}
\end{figure}

Instead, we approached the problem by minimizing a figure of merit 
$\Delta = \Sigma_i (T_{dyn,i} - T_{morph,i})^2$ for various classifiers. Here $T_{dyn}$ is a proposed 
kinematic classifier based on some combination of the three key variables, $v$, $\sigma$
and $\epsilon$, and $T_{morph}$ is the HST-based type, E (T$=0$), E/S0
(T$=1$) or S0 (T$=2$). In order to use the visual morphologies to
calibrate our kinematic classifiers in this way, we must ensure that our
visual classification is unbiased. In some local samples, a deficit of 
face-on S0s seems to indicate that many are misidentified as 
ellipticals \citep{jf94}. In Figure~\ref{ellip}, we display cumulative
histograms of ellipticity for Es and S0s in our combined cluster and field
sample of 67 sources (in red). It is clear from the figure that the 
distributions of ellipticities in our sample are consistent with the
grey model curves overplotted, which indicate a Gaussian distribution of 
ellipticities with randomly distributed inclinations. As such,
our sample is an appropriately unbiased calibrator for our kinematic classifiers.
After minimizing $\Delta$ for several candidate indicators, we found
that the combinations $v/\,(1-\epsilon)$ vs $\sigma$ 
and $v/\,\sigma$ vs $\epsilon$ were the most successful, both yielding
a higher success rate than the canonical $v/\sigma^*$. 

The former criterion yields optimal separation of Es and S0s when we
classify galaxies with $v/\,(1-\epsilon)>85\pm5$ km s$^{-1}$ as S0s. 
We note that this choice of kinematic classifier does not strictly 
require a measurement of $\sigma$, and so can be applied to observations
where $\sigma$ is unavailable.
Overall, the criterion recovers $63\%\pm3\%$ of the
35 S0s and $76\% \pm^8_3\%$ of the 29 Es in our sample (Figure~3,
bottom). Given there is inevitably some morphological
misclassification at the 10-20\% level \citep{ellis97,tt03}, 
this seems an adequate success rate. Specifically, all but two of the morphological S0s below 
the 85 km s$^{-1}$ limit display no rotation, i.e. they are consistent with 
$v<$30 km s$^{-1}$, and those morphological Es above the limit are mostly 
rotating with $v\simeq$60-180 km s$^{-1}$. Of course, the most
critical test of this discriminator is not its success rate at
identifying individual galaxies, but rather how well it estimates the
overall fraction of S0s. Morphologically, our sample contains 
43$\%\pm8\%$ ellipticals and 52$\%\pm8\%$ S0s, with the remaining 5$\%$
ambiguous E/S0s; note that we do not yet include any accounting for
spiral galaxies. The $v/\,(1-\epsilon)$ discriminator yields an S0 fraction of
43$\%\pm5\%$, which is consistent within the errors of the visually classified proportion.

We then considered a criterion that exploits $\sigma$ as well. While the
previous criterion was practical and effective, it was solely concerned with
detection of rotation for a given shape. The addition of $\sigma$ introduces
a measure of pressure support, and thus should be a more robust indicator of
the presence of a bulge. Specifically, if we define
ellipticals to have $v/\,\sigma<$0.50 $\pm$0.03 and $\epsilon<$0.3,
a somewhat higher degree of success is revealed: we correctly identify
80$\%\pm2\%$ and 79$\%\pm2\%$ of S0s and ellipticals, respectively
(Figure~3, top). Likewise, the
predicted overall fraction of S0s--51$\%\pm3\%$--is in excellent agreement with its
morphological equivalent.

Naturally, once we calibrate our kinematic classifiers against our
$z\sim0.5$ sample, it is essential that we validate its performance against an
independent set of data. As our goal is to develop a morphological
discriminator that is more redshift-invariant than visual
classification, it makes sense to choose a comparison sample at a much
different redshift. Accordingly, we independently 
applied both kinematic classifiers to a local sample of 35 Es and S0s in the
Coma cluster \citep{mehlert03}, supplemented by an additional 11
mostly elliptical galaxies from \citet{bn90}. The distribution
of ellipticities of S0s in this sample does appear to suffer from the
bias identified by \citet{jf94} (Figure~\ref{ellip}). However, since we are 
not directly calibrating our kinematic indicators on this local comparison 
sample, the bias in the sample will not prevent us from evaluating 
the performance of our indicators at low redshift. With our small
sample size, we calculate that the bias reflected in
Figure~\ref{ellip} implies that only 1--2 additional galaxies with
incorrect visual morphologies are present in the sample, which is
small compared to other sources of uncertainty.


Importantly, Mehlert et al. published full rotation curves for their galaxies,
allowing us to re-measure velocities in a manner similar to that adopted
in $\S$3.  For each of their galaxies, we fit a straight line to data points within 
radius $<0.25R_e$, thus simulating the radial extent of our
curves. Such a truncated fit generally yields velocities about
40$\%$ of the velocity revealed by the full rotation curves of
the local galaxies, and we expect that our $z\sim0.5$ sample similarly underestimates rotation.
Such underestimated velocities may contribute to our
misidentification rate, suggesting that deeper observations at
$z\ge0.5$ may improve the accuracy of our kinematic classification. 

\begin{figure}
\includegraphics[width=\columnwidth]{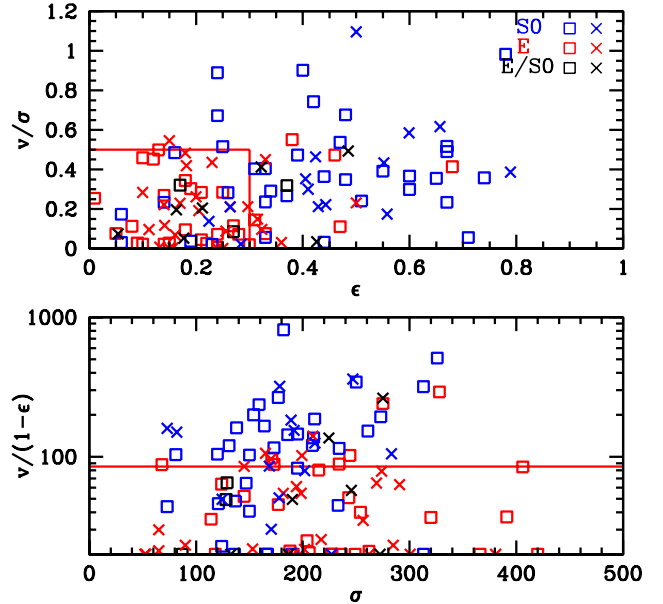}
\caption{The efficacy of two kinematic classifiers in recovering morphological
types in the distant cluster sample. In both cases, symbol colors denote the visually
determined morphologies. Squares refer to distant cluster galaxies and crosses to
those in the local sample. 
(Top) $v/\,\sigma$ vs $\epsilon$: adopting a limit of $v/\sigma>0.50$ and $\epsilon>0.3$ 
correctly identifies $>79\%$ of Es and S0s.  (Bottom) $v/\,(1-\epsilon)$ vs $\sigma$: adopting a rotational criterion
of $v/\,(1-\epsilon)>$85 km s$^{-1}$ is an accurate substitute,
correctly identifying 63$\%$ of S0s and 76$\%$ of Es.}
\label{fig: dyn_classification}      
\end{figure}

As shown in Figure~3, the recovery is equally successful (65-80\% accuracy 
for individual galaxies) for both kinematic classifiers. They likewise reliably 
recover the correct morphological mix. The $v/(1-\epsilon)$
discriminator predicts that the local sample consists of $30\%\pm3\%$
S0s, identical to the visually-determined fraction, while the other 
discriminator predicts $37\%\pm3\%$. This 
demonstrates no obvious redshift dependence in the classification, at least 
within the errors of the small available samples. In Table~2, we
list the success rates achieved for each classifier on both local
ans $z\sim0.5$ data, as well as the calculated S0 fractions from each method.

\begin{deluxetable*}{ccccccccc}
  \tablewidth{0pt}
  \tablecaption{Success rates of kinematic classifiers}
  \tablenum{2}
  \tabletypesize{\scriptsize}

  \tablehead{
\colhead{Sample} & \multicolumn{3}{c}{$v/\sigma$ vs $\epsilon$}
& \colhead{} & \multicolumn{3}{c}{$v/(1-\epsilon)$ vs $\sigma$} & \colhead{Visual
  Class.} \\  
\cline{2-4} \cline{6-8} \\ 
\colhead{} & \colhead{E} & \colhead{S0} & \colhead{$f_{S0}$} &
\colhead{} & \colhead{E}
& \colhead{S0} & \colhead{$f_{S0}$} & \colhead{$f_{S0}$}}

  \startdata
 z$\sim0.5$ & $80\pm2\%$ & $79\pm2\%$ & $51\pm3\%$ & & $63\pm3\%$ & $76\pm8_2\%$ & $43\pm5\%$ & $52\pm8\%$\\
 z$\sim0.0$ & $79\pm3\%$ & $76\pm3\%$ & $37\pm3\%$ & & $64\pm3\%$ & $80\pm2\%$ &   $30\pm3\%$ & $30\pm8\%$ \\

\enddata

\tablecomments{For each kinematic discriminator, we list under heading E (S0) the fraction of visually-determined
Es (S0s) where the kinematic discriminator yields the same class. Under $f_{S0}$, we indicate the overall
fraction of S0s predicted by applying each kinematic classifier, in comparison to the visually-determined 
$f_{S0}$ at right.}

\end{deluxetable*}

Locally, it has been shown that a simple cut at $\epsilon=0.3$
effectively identifies $\sim70\%$ of S0s and an even higher fraction
of Es \citep{jf94}, and in our
$z\sim0.5$ sample the success rate is only slightly worse ($\sim70\%$ of
both Es and S0s). So what, then, have we gained by
adding kinematic information to this mix?  The advantages are
two-fold. First, while $\epsilon$ may be an accurate predictor of
morphology locally, such a relation is known to break down toward
higher redshift, where elongated galaxies do not necessarily reflect
ordered rotation \citep{erb04}. Velocities and dispersions therefore
become essential in separating rotating from non-rotating galaxies as 
we extend our current method of E--S0 discrimination to higher
redshift.  Being sensitive to the dynamical structure, we can also hope to 
be less influenced by band-shifting or recent
star formation, which could influence the measurement of ellipticity.
Second, applying two different kinematic indicators, such as
those presented here, adds confidence to the determination of the
morphological mix, so long as the two indicators yield consistent
measures of the S0 fraction. Overall, then, ellipticity is most
usefully combined with measures of rotation and velocity dispersion 
in order to assess of the morphologies of galaxies, while minimizing
redshift-dependent effects.

\section{EVOLUTION OF KINEMATICALLY-DEFINED S0S}

To demonstrate the utility of kinematic information in studying the
evolution in the abundance of S0s, we now apply the kinematic
classifiers to re-visit the evolving morphology-density relation,
tagging each galaxy according to its local environmental density,
$\Sigma$, as defined in \citet{tt03}.
In constructing the fraction of S0s, $f_{S0}$, at $z\sim0.5$, we assume
we have representatively sampled the available population. To improve
statistics in this preliminary comparison, we combine data from both clusters
and contrast this with estimates in the field, for which $\Sigma$ estimates
have been determined following the procedures discussed in \citet{tt05a}.
To estimate $f_{S0}$, we first recalculate the robust visual determination of
$f_{E+S0}$ from \citet{tt03} into two density bins across
the cluster fields, and also adopt the field value from that paper. We
then calculate $f_{S0}$ by counting the number of kinematically-determined ellipticals ($N_E$)
and S0s ($N_{S0}$) in each density bin and the field, such that 
$f_{S0}=N_{S0}/(N_E+N_{S0})*f_{E+S0}$. 

Figure~4 illustrates the result. We find a fairly low fraction of S0s ($f_{S0} \simeq$28$\pm$6\%) 
at all cluster densities, except that our sample is not yet sufficiently large to
reliably probe the core regions. The local S0 fraction at comparable densities
is about $f_{S0}$=50\% \citep{d80} so the kinematic data strongly support
earlier contentions \citep{d97,postman05} that there is
a substantial decline in the S0 fraction in clusters. Interestingly, the fraction
in the field is even lower, $16\% \pm 5\%$, although clearly non-zero. Larger
samples would confirm these trends.

\begin{figure}
\includegraphics[width=\columnwidth]{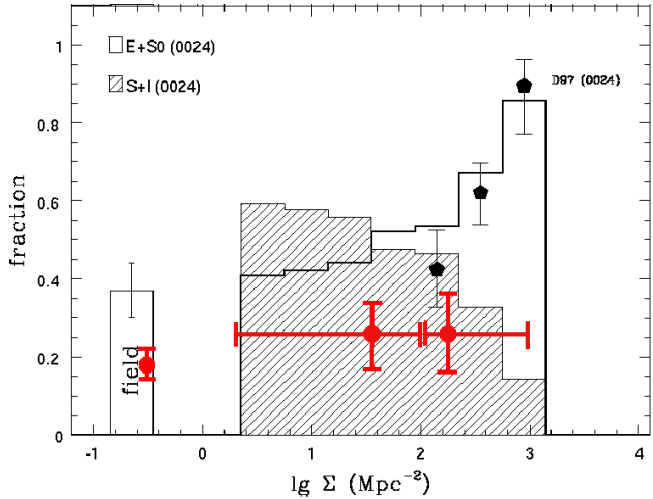}
\caption{Morphology-density relation at $z\simeq$0.5, adapted from
  Treu et al. (2003). Red points indicate the S0 fraction as
  determined from our kinematic indicators, with error bars
  reflecting the disagreement between discriminators, plus
  Poisson uncertainty. Histograms show E+S0
  and Spiral+Irregular fraction from Treu et al. (2003), as indicated.
  Black points are E+S0 fraction from Dressler et al. (1997).}
\label{fig:TS_S0}
\end{figure}

Recently, \citet{vv06} have also considered
the utility of resolved kinematic data of distant cluster galaxies. Their
analysis is more concerned with establishing the fraction of rotating
spheroidals rather than in separating Es and S0s.  Based on a much smaller
sample, they find a slightly {\it larger} fraction of rotating Es than 
observed locally. We note they use the quantity $v/\,\sigma^*$ which we
found to be ineffective for reasons discussed earlier.

Given the efficiency of multi-object spectrographs such as DEIMOS, it
is interesting to consider the prospects of our method for tracing the
full evolutionary history of S0s in various density regimes over 0$<z<$1,
for example in order to verify or otherwise the scenario put forward by
\citet{smith05} and \citet{postman05}. Visual discrimination between early
and late-type galaxies is more reliable than
E--S0 separation, even up to $z>1.0$ \citep{postman05}, and so our
reliance on the visually-determined $f_{E+S0}$ poses no barrier to 
extending this method up to $z\sim1$. 

The key challenge to extending our method to higher redshift is the loss of
spatial resolution that occurs as one observes galaxies at smaller apparent
sizes, with a relatively fixed ground-based seeing limit. At
$z\sim0.5$, our typical seeing of $0\farcs7$ FWHM
corresponds to 4.3~kpc, compared to a 6--8~kpc typical extent
of our rotation curves. At $z=1$, 
the same seeing disk covers a physical diameter of 5.6~kpc, 30\%
larger. Considering our requirement to have at least three independent spatial bins, and 
assuming similar signal to noise spectra, a similar study at $z\sim1$
may be limited to galaxies with spatial extent $\gtrsim8$~kpc; only
about 25\% of the $z\sim0.5$ sample meet this requirement.

In fact, the situation is not so dire, as we are most interested in
detecting only whether a galaxy is rotating or not, and so blurring of
the velocity gradient due to seeing is only important if the velocity
is smeared toward zero. To test our ability to measure $z\sim1$ rotation
curves, we have convolved our $z\sim0.5$ curves with a Gaussian kernel
to simulate the seeing at $z=1$. Remeasuring velocities on these blurred
curves, we find that our kinematic classification scheme predicts the
same morphology as the unblurred curve in 96$\%$ of the objects.
We thus conclude that kinematic discrimination between Es
and S0s can be a powerful tool for tracing the presumed buildup of S0
galaxies even up to $z>1$, and will provide an
additional and more fundamental discriminator at these high redshift.

\section{Conclusions}

We examine the potential of using resolved kinematics of distant galaxies
to separate, independently of morphological data, elliptical and lenticular
galaxies. Applying various criteria to a sample of 44 cluster galaxies and
23 field galaxies at $z\simeq$0.5, we find promising prospects. Using
the morphological classification as a starting point, we recover the
morphological mix to within 10\% accuracy using various
combinations of the stellar rotational
velocity, velocity dispersion and ellipticity. We test the utility of our classifiers
on local data and use them to establish the first kinematically-based evidence
for a declining fraction of S0s with redshift across a wide range of
densities.  

\begin{acknowledgements}

Faint object spectroscopy at with DEIMOS at Keck Observatory is made
possible with the efforts of P. Amico, S. Faber, 
and G. Wirth. The analysis pipeline for reducing DEIMOS data was 
developed at UC Berkeley with support from NSF grant AST-0071048. 
RSE acknowledges financial support from NSF grant 
AST-0307859 and STScI grants HST-GO-08559.01-A and 
HST-GO-09836.01-A.

\end{acknowledgements}

\clearpage

\begin{deluxetable}{crrcccccc}
  \tablewidth{0pt}
  \tablecaption{Measurements of Cl~0024+17, MS~0451-03
   and field galaxies}
  \tablenum{1}
  \tabletypesize{\tiny}

  \tablehead{\colhead{ID} & \colhead{RA} & \colhead{DEC} & \colhead{z}
  & \colhead{v}& \colhead{$\sigma$}&
  \colhead{$\epsilon$} & \colhead{Visual} & \colhead{$\delta$}
  \\ \colhead{} & \colhead{($^\circ$)} & \colhead{($^\circ$)} &
  \colhead{} & \colhead{(km s$^{-1}$)}& \colhead({km s$^{-1}$)} &
  \colhead{} & \colhead{Morph} & \colhead{$(^\circ)$}}
  \startdata
C1 & 6.682304 & 17.138241&  0.397&$137 \pm 9$ & $211 \pm 11$ & 0.24 & S0 & 15 \\
C2 & 6.726579 & 17.140829&  0.393&$65 \pm 9$ & $234 \pm 22$ & 0.25 & E & 12 \\
C3 & 6.536877 & 17.165190&  0.397&$86 \pm 20$ & $186 \pm 24$ & 0.39 & S0 & 12 \\
C4 & 6.649991 & 17.162821&  0.391&$< 60$ & $262 \pm 25$ & 0.05 & E & 26 \\
C5 & 6.643103 & 17.172791&  0.386&$48 \pm 23$ & $215 \pm 22$ & 0.19 & E & 42 \\
C6 & 6.782351 & 17.180410&  0.397&$101 \pm 10$ & $209 \pm 16$ & 0.16 & S0 & 5 \\
C7 & 6.631488 & 17.286880&  0.393&$141 \pm 10$ & $182 \pm 20$ & 0.78 & S0 & 38 \\
C8 & 6.514315 & 17.317190&  0.395&$< 29$ & $227 \pm 29$ & 0.21 & E & 9 \\
C9 & 6.631382 & 17.101040&  0.398&$41 \pm 18$ & $128 \pm 42$ & 0.17 & E/S0 & 2 \\
C10 &6.640234 & 17.158600&  0.392& $53 \pm 15$ & $195 \pm 22$ & 0.60 & S0 & 25 \\
C11 &6.660044 & 17.166290&  0.398& $41 \pm 8$ & $129 \pm 22$ & 0.37 & E/S0 & 4 \\
C12 &6.645772 & 17.172569&  0.388& $62 \pm 13$ & $173 \pm 18$ & 0.10 & E & 39 \\
C13 &6.630862 & 17.174200&  0.395& $23 \pm 8$ & $137 \pm 21$ & 0.33 & S0 & 44 \\
C14 &6.630893 & 17.182131&  0.396& $ < 58$ & $391 \pm 51$ & 0.24 & E & 45 \\
C15 &6.613021 & 17.207199&  0.398& $23 \pm 19$ & $204 \pm 30$ & 0.08 & E & 3 \\
C16 &6.595411 & 17.208710&  0.397& $33 \pm 15$ & $150 \pm 25$ & 0.14 & S0 & 20 \\
C17 &6.589549 & 17.236601&  0.396& $63 \pm 9$ & $150 \pm 16$ & 0.25 & S0 & 35 \\
C18 &6.741905 & 17.266029&  0.390& $7 \pm 4$ & $87 \pm 19$ & 0.27 & E/S0 & 20 \\
C19 &6.629107 & 17.285839&  0.396& $46 \pm 8$ & $120 \pm 18$ & 0.55 & S0 & 12 \\
C20 &6.499555 & 17.336269&  0.395& $39 \pm 11$ & $131 \pm 15$ & 0.60 & S0 & 36 \\
C21 &73.520592 & -3.096000&0.510& $153 \pm 34$ & $177 \pm 33$ & 0.40 & S0 & 16 \\
C22 &73.520485 & -3.000458&0.542& $30 \pm 24$ & $320 \pm 21$ & 0.18 & E & 6 \\
C23 &73.523880 & -3.011362&0.539& $34 \pm 21$ & $243 \pm 17$ & 0.31 & E & 14 \\
C24 &73.552063 & -3.018945&0.532& $130 \pm 20$ & $275 \pm 18$ & 0.46 & E & 0 \\
C25 &73.566452 & -3.101381&0.550& $< 81$ & $406 \pm 63$ & 0.47 & E & 30 \\
C26 &73.439285 & -3.137721&0.545& $< 37$ & $420 \pm 82$ & 0.30 & E & 31 \\
C27 &73.442268 & -2.900079&0.552& $< 81$ & $254 \pm 23$ & 0.27 & E & 41 \\
C28 &73.530807 & -3.091694&0.546& $93 \pm 44$ & $313 \pm 36$ & 0.65 & S0 & 33 \\
C29 &73.496529 & -2.947528&0.548& $< 38$ & $227 \pm 17$ & 0.22 & E & 1 \\
C30 &73.512939 & -2.932262&0.544& $145 \pm 39$ & $326 \pm 39$ & 0.67 & S0 & 30 \\
C31 &73.501221 & -2.982960&0.533& $51 \pm 30$ & $234 \pm 27$ & 0.51 & S0 & 25 \\
C32 &73.559761 & -3.025668&0.540& $< 26$ & $190 \pm 26$ & 0.23 & S0 & 6 \\
C33 &73.545235 & -3.014432&0.539& $< 50$ & $366 \pm 52$ & 0.24 & E & 16 \\
C34 &73.540581 & -3.004315&0.545& $< 35$ & $188 \pm 18$ & 0.33 & E & 9 \\
C35 &73.545334 & -3.026837&0.535& $< 23$ & $136 \pm 10$ & 0.33 & S0 & 23 \\
C36 &73.541374 & -3.024953&0.539& $66 \pm 27$ & $210 \pm 17$ & 0.44 & S0 & 30 \\
C37 &73.558983 & -2.991541&0.532& $< 41$ & $249 \pm 40$ & 0.10 & E & 26 \\
C38 &73.579918 & -2.990385&0.531& $21 \pm 20$ & $68 \pm 24$ & 0.68 & E & 41 \\
C39 &73.601814 & -3.064165&0.546& $78 \pm 27$ & $250 \pm 24$ & 0.74 & S0 & 29 \\
C40 &73.600838 & -3.060720&0.536& $42 \pm 19$ & $177 \pm 14$ & 0.01 & E & 21 \\
C41 &73.690582 & -3.034126&0.543& $151 \pm 43$ & $328 \pm 25$ & 0.38 & E & 33 \\
C42 &73.664803 & -3.084279&0.540& $57 \pm 31$ & $273 \pm 15$ & 0.67 & S0 & 27 \\
C43 &73.555710 & -3.012476&0.532& $< 41$ & $233 \pm$ 46 & 0.71 & S0 & 0 \\
C44 &73.535393 & -3.009354&0.526& $78 \pm 15$ & $159 \pm$ 20 & 0.67 & S0 & 0 \\
F1 & 189.339612 & 62.226372&0.48&$81 \pm 25$ & $244 \pm 18$ & 0.18 & E & 15 \\
F2 & 189.374580 & 62.216985&0.51&$< 24$ & $191 \pm 20$ & 0.14 & E & 17 \\
F3 & 189.488759 & 62.263352&0.46&$< 20$ & $236 \pm 20$ & 0.15 & E & 40 \\
F4 & 189.220348 & 62.245666&0.32&$41 \pm 25$ & $145 \pm 9$ & 0.21 & E & 6 \\
F5 & 189.200321 & 62.219268&0.47&$56 \pm 13$ & $124 \pm 13$ & 0.12 & E & 2 \\
F6 & 189.252330 & 62.209726&0.56&$< 40$ & $118 \pm 13$ & 0.09 & E & 27 \\
F7 & 189.013560 & 62.186461&0.64&$24 \pm 17$ & $114 \pm 25$ & 0.14 & E & 39 \\
F8 & 189.073642 & 62.229072&0.53&$85 \pm 31$ & $171 \pm 12$ & 0.13 & E & 5 \\
F9 & 189.308295 & 62.343580&0.53&$< 35$ & $206 \pm 23$ & 0.28 & E & 19 \\
F10 &189.359264 & 62.229837&0.47& $104 \pm 16$ & $154 \pm 16$ & 0.48 & S0 & 2 \\
F11 &189.359027 & 62.234295&0.48& $< 41$ & $313 \pm 45$ & 0.44 & S0 & 2 \\
F12 &189.434130 & 62.232987&0.51& $25 \pm 17$ & $73 \pm 8$ & 0.33 & S0 & 32 \\
F13 &189.472041 & 62.248390&0.51& $114 \pm 10$ & $138 \pm 10$ & 0.24 & S0 & 22 \\
F14 &189.469027 & 62.247062&0.51& $88 \pm 15$ & $164 \pm 17$ & 0.47 & S0 & 1 \\
F15 &189.477354 & 62.257551&0.46& $31 \pm 10$ & $121 \pm 13$ & 0.26 & S0 & 25 \\
F16 &189.262189 & 62.239978&0.51& $< 120$ & $147 \pm 23$ & 0.34 & S0 & 11 \\
F17 &189.172782 & 62.234202&0.56& $< 19$ & $167 \pm 14$ & 0.19 & S0 & 13 \\
F18 &189.360790 & 62.287180&0.56& $< 13$ & $166 \pm 13$ & 0.06 & S0 & 11 \\
F19 &189.142570 & 62.242571&0.52& $52 \pm 49$ & $195 \pm 27$ & 0.37 & S0 & 5 \\
F20 &189.095001 & 62.216703&0.47& $58 \pm 17$ & $173 \pm 22$ & 0.48 & S0 & 16 \\
F21 &189.063660 & 62.206127&0.32& $46 \pm 9$ & $81 \pm 15$ & 0.42 & S0 & 40 \\
F22 &189.158314 & 62.291520&0.56& $74 \pm 19$ & $261 \pm 19$ & 0.31 & S0 & 45 \\
F23 &189.147118 & 62.186147&0.41& $21 \pm 10$ & $124 \pm 10$ & 0.06 & S0 & 12 \\
\enddata
\tablecomments{Galaxy IDs are designated with C for cluster members and
  F for field galaxies. Velocities are listed uncorrected for slit
  misalignment. Upper limits on velocities are 1$\sigma$.}
\end{deluxetable}

\end{document}